\documentclass[flushrt,preprint]{aastex}
\usepackage{amssymb}
\usepackage{amsmath}
\usepackage{graphicx}
\usepackage{psfrag}
\usepackage{dcolumn}
\newcolumntype{d}[1]{D{.}{.}{#1}}
\begin{document}
\psfrag{aaD}{$a D$}
\psfrag{aaoD}{$(1-a)D$}
\psfrag{ph}{$\phi$}
\psfrag{dt}{$\Delta\theta$}
\psfrag{pt}{$\pi-\Delta\theta$}
\psfrag{ch}{$\chi$}
\title{Inferences from the Distributions of Fast Radio Burst Pulse Widths,
Dispersion Measures and Fluences}
\shorttitle{FRB Pulse Widths}
\shortauthors{Katz}
\author{J. I. Katz\altaffilmark{}}
\affil{Department of Physics and McDonnell Center for the Space Sciences}
\affil{Washington University, St. Louis, Mo. 63130}
\email{katz@wuphys.wustl.edu}
\begin{abstract}
The widths, dispersion measures, dispersion indices and fluences of Fast
Radio Bursts (FRB) impose coupled constraints that all models must satisfy.
The non-monotonic
dependence of burst widths (after deconvolution of instrumental effects) on
dispersion measure excludes the intergalactic medium as the location of
scattering that broadens the FRB in time.  Temporal broadening far greater
than that of pulsars at similar high Galactic latitudes implies that
scattering occurs close to the sources, where high densities and strong
turbulence or heterogeneity are plausible.  FRB energetics are consistent
with supergiant pulses from young, fast, high-field pulsars at cosmological
distances.  The distribution of FRB dispersion measures is: 1) Inconsistent
with that of expanding clouds (such as SNR); 2) Inconsistent with
space-limited source populations (such as the local Supercluster); 3)
Consistent with intergalactic dispersion of a homogeneous source population
at cosmological distances.  Finally, the FRB $\log{N}$--$\log{S}$ relation
also indicates a cosmological distribution, aside from the anomalously
bright Lorimer burst.
\end{abstract}
\keywords{radio continuum: general --- intergalactic medium --- plasmas ---
scattering --- supernova remnants}
\maketitle
\section{Introduction}
Following an initial report of a fast radio burst by \cite{L07}, \cite{T13}
discovered four additional events whose large dispersion measures (DM $ >
500$ pc\,cm$^{-3}$) and high Galactic latitudes $\vert b \vert > 40^\circ$
indicated that their sources were at cosmological distances.  Subsequently,
more FRB were discovered (the present total is ten).  Their measured
parameters include fluence, dispersion measure (DM) and pulse widths.

Several FRB have dedispersed and deconvolved pulse widths $W_{1300}$ at 1300
MHz of several ms, while only upper limits were found for others.  Fitting
$W \propto \nu^\beta$ for FRB with measured $W$, their scattering indices
$\beta \approx -4$, consistent with theoretical values for multipath
propagation spreading in a plasma medium \citep{W72,LJ75,R77} and apparently
confirming that the burst widths are not instrumental
artefacts\footnote{Incomplete dedispersion of a burst might masquerade as
broadening with $\beta = -3$ because the propagation delay $\delta \Delta t
\propto (\delta \omega)^{-3}$ over a narrow frequency band $\Delta \omega$,
but $\beta = -3$ is outside the measured range of uncertainty of at least
one FRB \citep{T13}.}.  In homogeneous Kolmogorov turbulence $\beta = -4.4$,
consistent with all values of $\beta$ measured for FRB, but inconsistent
with many pulsar data \citep{KMNJM15}.

This paper explores the implications of the distributions of FRB scattering
widths, fluences and dispersion measures and the assumption that the
dispersion measures indicate cosmological distances.  Any explanation of
these observations must first account for two facts: (1) All FRB have
dispersion measures within a range of a factor of about three (two if the
Lorimer burst is not accepted as an FRB), implying a general property, not
an unusual circumstance such as a line of sight that happens to intersect a
rare dense cloud; (2) The dispersion index $\alpha$, defined by the
dispersive delay relation $\Delta t \propto \nu^\alpha$, is very close to
$-2$ and consistent with exactly $-2$ for every FRB, implying an upper bound
on the density of the dispersing plasma and (combined with the measured DM)
a lower bound on its size.  These facts are readily accounted for if most of
the dispersion is intergalactic, in which case the conventional assumptions
that most of the cosmic baryon density is ionized and homogeneous permits
the inference of the source redshifts $z$ from the observed DM.

Searches for FRB typically impose a minimum DM of 200 pc\,cm$^{-3}$ as
a filter against local interference and a maximum DM of 2000 pc\,cm$^{-3}$
because of pulse smearing in finite-width receiver filters, so that the
absence of observed FRB DM outside the range 200--2000 pc\,cm$^{-3}$ does not
imply their non-existence.  Galactic FRB at high latitude would be
essentially unobservable because of the low DM cutoff, unless they are
surrounded by dense ionized matter, and FRB with cosmological redshift $z \ge
1.4$ are not observable because of the high DM cutoff.

The frequency-dependent temporal broadening poses a different problem.  It
is difficult to attribute it to intergalactic space or to our Galaxy at the
high Galactic latitudes of eight of the ten reported FRB because much
lower upper bounds can be placed on the broadening of at least some Galactic
pulsars \citep{S04,HE07}.  This suggests
that the broadening originates near the sources of the FRB.  If their hosts
resemble the galaxies known to us, they would be expected to contribute
(unless fortuitously edge-on spirals) only a small fraction of either the
broadening or the dispersion.  Whatever the nature of the compact objects
making FRB, it is implausible that the properties of their host galaxies
differ by an order of magnitude from those of galaxies known to us.  This
suggests that the pulse broadening occurs in an immediate environment of
the FRB that may differ from the general interstellar medium of the host
galaxies (and of our own).

\begin{table}[h!]
\centering
\begin{tabular}{|c|c|c|r|c|l|}
\hline
FRB & DM$_E$ (pc\,cm$^{-3}$) & $b$ & $S$ (Jy\,ms) & $W_{1300}$ (ms) & Reference \\
\hline
010621 & {\it 213} & $-\phantom{0}4.0$ & $4.3\quad$ & $< 3\phantom{.0}$ & \cite{K12} \\
010724 & 350 & $-41.8$ & $150\phantom{.0}\quad$ & $6.2$ & \cite{L07} \\
121102 & {\it 369} & $-$\phantom{0}0.2 & $1.2\quad$ & $< 0.5$ & \cite{S14b} \\
120127 & 521 & $-66.2$ & $0.8\quad$ & $< 1.1$ & \cite{T13} \\
140514 & 528 & $-54.6$ & $1.3\quad$ & $1.9$ & \cite{P15a} \\
110626 & 677 & $-41.7$ & $0.9\quad$ & $< 1.4$ & \cite{T13} \\
010125 & 680 & $-20.0$ & $5.6\quad$ & $5\phantom{.0}$ & \cite{BSB14} \\
131104 & 710 & $-22.2$ & $2.3\quad$ & $4\phantom{.0}$ & \cite{R15} \\
110220 & 910 & $-54.7$ & $7.3\quad$ & $5.6$ & \cite{T13} \\
110703 & 1072 & $-59.0$ & $1.8\quad$ & $< 4.3$ & \cite{T13} \\
\hline
\end{tabular}
\caption{Summary of FRB data, ordered by estimated extra-Galactic dispersion
measures DM$_E$; for the two low-$b$ (Galactic latitude) bursts these may be
uncertain and are italicized.  Most data are from references in the table,
but fluences $S$ are from the re-evaluation by \cite{KP15}, except for the
``Lorimer burst'' (FRB 010724) and FRB 131104.  Pulse widths $W_{1300}$ are
from the references after dedispersion and deconvolution of the instrumental
response and scaled, if necessary, $\propto \nu^{-4}$ to 1300 MHz for
consistency with \cite{T13} (scalings are over modest frequency ratios and
insensitive to the scaling exponent).}
\label{datatable}
\end{table}

The Table summarizes the published data on FRB.
Section~\ref{wherenot} discusses the origin of the broadening of
some FRB during propagation.  Section~\ref{PSR} considers a specific FRB
model, supergiant pulses from fast, young, high-field pulsars.  This
leads to consideration of young SNR as the possible origin of FRB
dispersion, in which case FRB distances would not be cosmological.
Dispersion local to the FRB is discussed in Section~\ref{local}; SNR are
rejected on the basis of the distribution of dispersion measures, but
massive quasi-static clouds are not.  Section~\ref{Nsources} constrains the
number of active FRB sources, while Section~\ref{SNrate} compares this to
the supernova rate.  Section~\ref{logNlogS} uses the $\log{N}$--$\log{S}$
relation to constrain the space distribution of FRB.  The hypothesis of
cosmological distances inferred from the dispersion measures of FRB is
tested in Section~\ref{DMcosmo} and found to fit the distribution of
dispersion measures well.  Section~\ref{discuss} contains a summary
discussion.  Appendix~\ref{chirp} argues that FRB are unlikely to be chirped
radar pulses.
\section{Bounds Implied by Dispersion Indices}
\label{dindex}
The dispersion index $\alpha$, defined by the dispersion delay $\Delta t
\propto \nu^\alpha$, is a strong constraint on the density of the dispersing
plasma.  The tightest constraint is for FRB 140514 for which $\alpha =
-2.000 \pm 0.004$ \citep{P15a}; all measured values of $\alpha$ are
consistent with -2, though others have somewhat larger uncertainties.
Expansion of the dispersion relation for electromagnetic waves in a cold
(nonrelativistic) plasma in powers of $\omega_p^2/\omega^2 \ll 1$ yields
\citep{K14b}
\begin{equation}
\Delta t = \int\!{d\ell \over c}\,{1 \over 2}{\omega_p^2 \over \omega^2}
\left(1 + {3 \over 4}{\omega_p^2 \over \omega^2} + \cdots\right)
\end{equation}
and
\begin{equation}
\label{alpha}
\alpha \equiv {d \ln{\Delta t} \over d \ln{\omega}} = -2 -{3 \over 2}
{\omega_p^2 \over \omega^2} + \cdots = -2 -{6 \pi n_e e^2 \over 
m_e \omega^2} + \cdots.
\end{equation}
If dispersion occurs at a significant redshift then $\omega$ is the
frequency in the region in which the dispersion takes place (blue-shifted
from the frequency of observation) and the plasma frequency $\omega_p$
refers to the electron density there \citep{DGBB15}.

In order to constrain $n_e$ in the scattering region we must allow for
the fact that it contributes only $\mathrm{DM_{scatt}}$ to the
(extra-Galactic) dispersion of the pulse.  The remainder, perhaps nearly
all, is attributed to intergalactic propagation, for which the $\omega_p^2/
\omega^2$ and higher terms in (\ref{alpha}) are negligible.  From the
observed bounds on $\alpha$, (\ref{alpha}) yields
\begin{equation}
\label{DMlocal}
{\omega_p^2 \over \omega^2} {\mathrm{DM_{scatt} \over DM}} \le {2 \over 3}
\max{(-\alpha - 2)} \lessapprox 0.003,
\end{equation}
where $\max{(-\alpha - 2)} \approx 0.004$ is the observed upper bound on
$-\alpha - 2$ for FRB 140514 \citep{P15a}.  Using 
$\omega_p^2 = 4 \pi n_e e^2/m_e$, (\ref{alpha}) and (\ref{DMlocal}),
\begin{equation}
\label{nebound}
n_e \lesssim {1 \over 6 \pi} {m_e \omega^2 \over e^2} \max{(-\alpha -2)}
\approx 1 \times 10^8\ \mathrm{cm}^{-3}.
\end{equation}
The size $R$ of the dispersing region is bounded:
\begin{equation}
R = {\mathrm{DM} \over n_e} \gtrsim 2 \times 10^{13}\ \mathrm{cm}.
\end{equation}
This temperature-independent limit excludes models that attribute the
dispersion to the immediate environment of a star.

Analogous but temperature-dependent limits are implied by the requirement
that the dispersing cloud be transparent to inverse bremsstrahlung
absorption.  The inverse bremsstrahlung limits scale $\propto T^{-3/2}$ and
are relaxed if absorption of an energetic pulse heats the plasma.
\section{Pulse Widths}
\label{wherenot}
We approximate the burst propagation as that in a flat static (Euclidean)
universe.  For the estimated redshifts $z \le 0.96$ inferred from the
dispersion measures by attributing the dispersion to intergalactic plasma
with conventional cosmological parameters (the overwhelming majority of the
baryon density distributed homogeneously in intergalactic space \citep{M15};
possible inhomogeneity is discussed by \cite{Z14}), this only introduces an
error of a factor ${\cal O}(1)$, less than other uncertainties.

\begin{figure}[!h]
\centering
\includegraphics[width=2.5in]{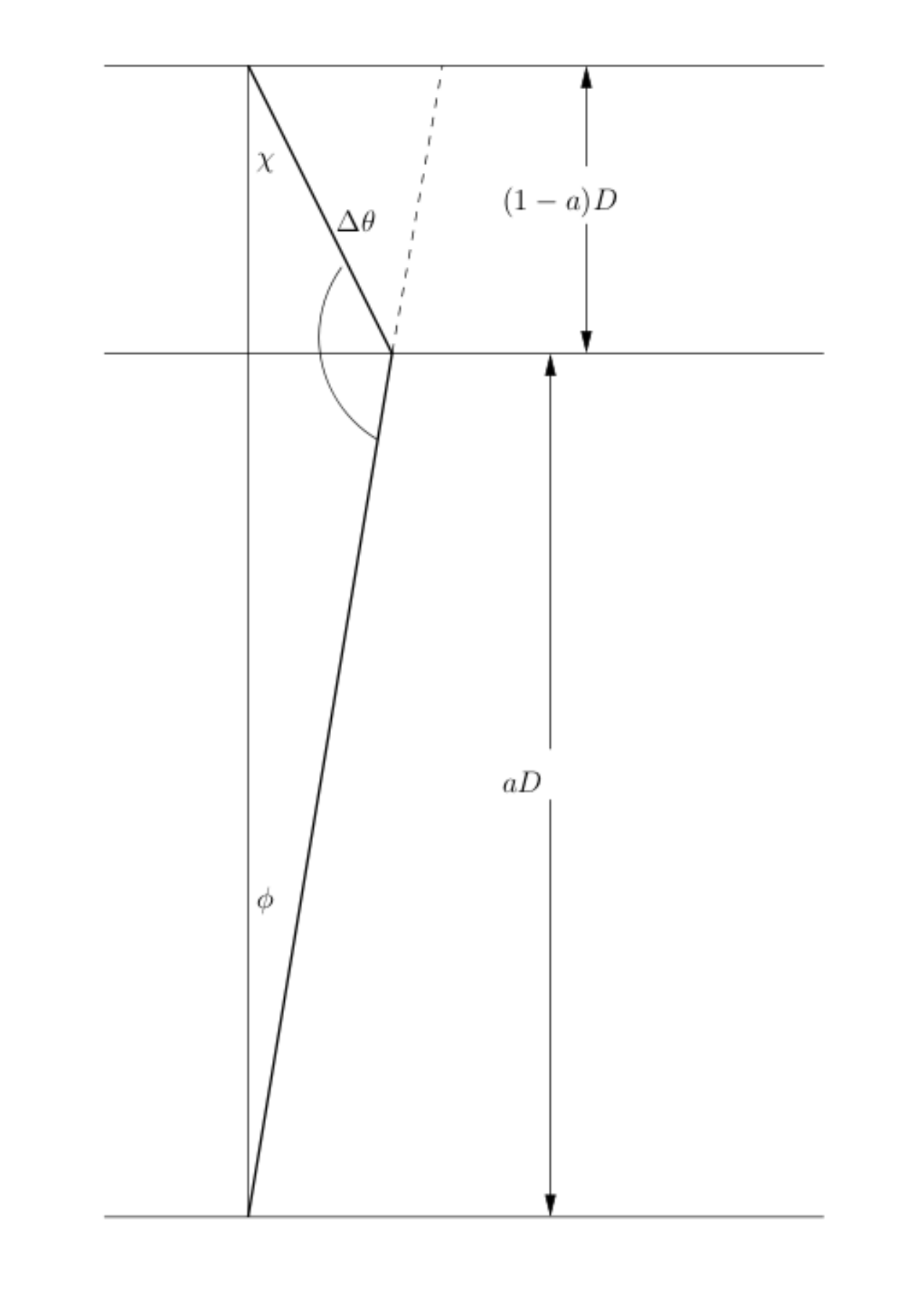}
\caption{\label{pathfig}Path of scattered radiation \citep{W72,Ku14}}
\label{propagation}
\end{figure}

Following the classic model of \cite{W72,LJ75,R77}, we approximate the
propagation paths as produced by a single scattering at a distance $aD$ from
us and $(1 - a)D$ from the source.  If the scattering angle $\Delta \theta
\ll 1$ then the angles $\phi \approx (1 - a) \Delta \theta$ and $\chi
\approx a \Delta \theta$; the geometry is shown in Fig.~\ref{propagation}.

We assume that the origin of the pulse width $W$ is dispersion in
propagation path lengths.  The incremental delay attributable to scattering
by an angle $\Delta \theta$ \citep{W72,Ku14} is
\begin{equation}
\label{delay}
W \approx {D \over 2 c} (\Delta \theta)^2 a (1 - a).
\end{equation}
Then
\begin{equation}
\label{Dtheta}
\Delta \theta \approx \sqrt{2 c W \over D a (1-a)} \gtrsim \sqrt{8 c W \over
D} \sim 5 \times 10^{-10},
\end{equation}
for all FRB with measured $W$.  The minimum value is obtained for $a = 1/2$.
The angular width of the received radiation
\begin{equation}
\label{phi}
\phi \approx (1 - a) \Delta\theta = \sqrt{{2 c W \over D}
\left({1-a \over a}\right)}.
\end{equation}

Because of its low density, the intergalactic medium is unlikely to be the
location of scattering.  The Table shows that the pulse widths $W_{1300}$ do
not vary monotonically with dispersion measure, in contrast to expectations
for intergalactic scattering \citep{MK13}.  In addition, \cite{LG14} have
shown that Kolmogorov turbulence in the intergalactic medium cannot explain
the pulse broadening because the implied dissipation of turbulent energy
would be excessive and the turbulence would decay rapidly (this argument
does not apply to quasi-static isobaric heterogeneities such as the
multi-phase structure of the interstellar medium, and its possible 
intergalactic analogues).

If a scattering screen is close to the source ($1 - a \ll 1$) or to the
observer ($a \ll 1$) then $W \propto D \min{(a,1-a)}$.  The pulse broadening
produced by Galactic scattering ($a \ll 1$) is comparable to that of a
Galactic pulsar in the same direction because the screen's distance $aD$ is
similar, even though $D$ may be seven orders of magnitude greater for the
FRB, while $1 - a = {\cal O}(1)$ for both the Galactic pulsar and the
distant FRB. 

Galactic pulsars at high $b$ show orders of magnitude less pulse broadening
than FRB \citep{KMNJM15}.  Even at low $b$, the nanoshots of the Crab pulsar
\citep{HE07} imply, assuming $\nu^{-4}$ scaling from 9.25 GHz, broadening
$\lesssim 1.0\,\mu$s at 1300 MHz and those of PSR B1937+21, scaling only
from 1.65 GHz (hence insensitive to the assumed scaling relation),
broadening $\lesssim 40\,$ns \citep{S04}.  Unless these are the consequence
of focusing, with the same propagation time for all focused paths by
Fermat's principle, they indicate that Galactic scattering cannot contribute
significantly to the pulse widths of FRB.

Excluding intergalactic and Galactic scattering as the origin of the pulse
broadening implies it takes place near the FRB sources.  $\Delta\theta$ must
be orders of magnitude larger than the numerical value in Eq.~\ref{Dtheta}
because $1-a \ll 1$.  We know nothing of the host galaxies of FRB, but
galaxies at $z \le 1$ roughly resemble local galaxies.  By the preceding
argument, their general interstellar media cannot contribute a major
fraction of the pulse broadening (or of the dispersion measure) unless all
the five (four if the Lorimer burst is excluded) FRB with reported pulse
broadening are found in spirals viewed edge-on, a statistical improbability.

The immediate environment of the FRB must be much more strongly scattering
than ordinary interstellar media, indicating a causal relation with the FRB
itself.  FRB are likely to be young objects because compact, strongly
scattering, gas clouds would quickly dissipate (Section~\ref{SNR}) if
unbound.

\section{Pulsar Super-Pulses?}
\label{PSR}
There is an obvious resemblance between FRB and the giant pulses of some
radio pulsars.  Like radio pulsars, FRB are coherent emitters 
\citep{L07,T13,K14a}
with very high brightness temperatures.  Could FRB be the same phenomenon on
a more energetic scale, perhaps triggered by collapse of the neutron star
\citep{FR14}?

The high dispersion measures (and high Galactic latitudes) of FRB indicate
cosmological distances and much greater powers than inferred for Galactic
pulsars; \cite{T13} estimated, assuming isotropic emission, $P \gtrsim
{\cal O} (10^{42})$ erg/s for FRB 110220, where the lower limit assumes that
the observed pulse width is partly intrinsic and not entirely the result of
scattering.  This should be compared to the classic result for a
rotation-powered pulsar's spindown power (converted to radio emission with
an efficiency that is $\ll 1$ for all known radio pulsars)
\begin{equation}
\label{PSRpower}
P_{spindown} \approx {2 \mu^2 \Omega^4 \over 3 c^3} \approx 2 \times 10^{44}
\mu_{30}^2 \Omega_4^4\ \mathrm{erg/s},
\end{equation}
where $\mu_{30} \equiv \mu/10^{30}\,\text{gauss\,cm}^3$ is the scaled
magnetic dipole moment and $\Omega_4 \equiv \Omega/10^4\,s^{-1}$ is the
scaled rotation frequency.  This relation is exact for an oblique
($90^\circ$) rotator in vacuum and is believed to be approximately correct
for any angle between magnetic and rotational axes.  The angular
distribution of this power is unknown; dipole emission and winds from
spindown of aligned rotors are roughly isotropic, but how this power is
converted to coherent GHz radio emission, and its resulting angular
distribution, is not understood.

Eq.~\ref{PSRpower} indicates that a combination of high dipole moment and
fast spin would be required to explain FRB as rotation-powered pulsars.
Radio pulsars are known with such high values of $\mu$ and $\omega$, but not
in combination.  The combination, if it occurs, would lead to a short
spin-down time
\begin{equation}
\label{spindown}
t_{spindown} = {1 \over 2} {I \Omega^2 \over P_{spindown}} = {3 \over 4}
{I c^3 \over \mu^2 \Omega^2} \approx {10\ \mathrm{y} \over (\mu_{30}
\Omega_4)^2} \lesssim 10^3 \Omega_4^2\ \mathrm{y},
\end{equation}
where $I \sim 10^{45}$ g\,cm$^2$ is the moment of inertia and the last
inequality uses Eq.~\ref{PSRpower}.  For a neutron star spinning near
breakup and $P_{spindown}$ as large as required by a cosmological distance
for FRB 110220 \citep{T13}, $t_{spindown}$ could be as long as $\sim
10^3\,$y, but this would require implausibly efficient emission and optimal
choice of parameters.
\subsection{Soft Gamma Repeaters?}
A fundamental property of rotation-powered pulsar models is that their 
instantaneous power (including that into a particle wind that is believed to
dominate the energetics) is limited by Eq.~\ref{PSRpower}; there is no
intermediate energy reservoir between the rotation and accelerated particles
or radiated fields that could be drawn down in sudden bursts of higher
power.  Rotation-powered pulsar models are thus distinguished from
``magnetar'' models \citep{K82,TD92,M08} of soft gamma repeaters (SGR)
powered by dissipation of magnetic energy.  The upper limit to the rate of
energy release in a magnetar model is set by the poorly understood process
of magnetic reconnection, and might be as high as ${\cal O}(R_{NS}^2 B^2 c)
\sim 10^{46} \mu_{30}^2$ erg/s, as required to account for the December 27,
2004 outburst of SGR 1806$-$20, assuming isotropic emission, as would
be expected for a magnetically trapped pair plasma.

In SGR this power is thermalized, appearing as roughly black body emission
of soft gamma rays or hard X-rays, not as the high brightness radio
frequency emission or relativistic particle acceleration of rotation-powered
pulsars; thermalization may be inevitable at intensities $\gtrsim 10^{29}$
erg/cm$^2$s$^{-1}$ \citep{K96}.  Perhaps the same object might be both a SGR
and an FRB \citep{Ku14}, though at different times or at different places in
its magnetosphere.  For example, PSR J1745$-$2900 is observed both as an
anomalous X-ray pulsar (and hence is often termed a magnetar; \cite{M13})
and as a radio pulsar \citep{E13} emitting high-brightness nonthermal
radiation.

A Galactic FRB, if above the horizon, could have been detected in the far
sidelobes of any radio telescope observing in the L-band \citep{K14a} with
sufficient time resolution to detect a transient.  No such detections of the
known giant outbursts of Galactic SGR have been reported, but it is unclear
if any radio telescope was operating in a suitable mode, if the SGR were
above the horizon, or if such a signal would have been rejected as
interference.
\section{Dispersion Local to FRB?}
\label{local}
\subsection{Distribution of Dispersion Measures in Young SNR}
\label{SNR}
Some of the difficulties of the pulsar supergiant-pulse hypothesis would be
mitigated if the dispersion were produced in the pulsar's immediate
vicinity, so that the DM would not imply cosmological distances, reducing
the required energy \citep{PC15,CSP15}.  Rotation-powered pulsar models of
FRB imply short-lived, and hence young, sources, but the constraints of
Section~\ref{PSR} would be mitigated if they are cosmologically local.  If
so, a surrounding young SNR might be the chief source of dispersion
\citep{Ku14,CSP15}, possibly with a major Galactic contribution for the two
low-latitude FRB.  The local dispersion measure of a source at the center of
a spherical cloud of ionized gas of mass $M$ and radius $R$ is
\begin{equation}
\label{SNRDM}
\mathrm{DM_{local}} = 818 {M \over M_\odot} \left({R \over 0.1\,\mathrm{pc}}
\right)^{-2} f\ \mathrm{pc\,cm^{-3}},
\end{equation}
where $f = 1$ for a homogeneous sphere and $f = 1/3$ for a thin shell,
implying $R \sim 0.1\,$pc for a SNR, lost stellar envelope, {\it etc.\/}
providing much of the dispersion measure of an FRB.

As discussed in Section~\ref{wherenot}, this cloud must explain most of the
pulse broadening, whether or not it contributes most of the dispersion
measure.  If the customary Kolmogorov turbulence model of diffractive
scattering \citep{R77} is assumed, then
\begin{equation}
C_{n_e}^2 \approx 0.5 \left({W \over 5\,\mathrm{ms}}{1\,\mathrm{Gpc} \over D}
\right)^{5/6} \left({0.1\,\mathrm{pc} \over R}\right)\ \mathrm{m}^{-20/3}.
\end{equation}
This is comparable to values found for the more distant and highly dispersed
Galactic pulsars \citep{KMNJM15}, although the measured pulsar $C_{n_e}^2$
range over orders of magnitude.  \cite{LDKK13} have shown that the
scattering spectral indices of many pulsars are far from the Kolmogorov
model prediction of $-4.4$, and there is no {\it a priori\/} reason to expect
either this model or values of $C_{n_e}^2$ found for the Galactic
interstellar medium to be applicable to the environments of FRB.

The characteristic lifetime $T$ of an expanding cloud
\begin{equation}
\label{SNRlife}
T \approx {R \over V} \approx 30 {R \over 0.1\,\mathrm{pc}} {3000\,
\mathrm{km/s} \over V}\ \mathrm{y} \approx 30 \sqrt{{f \over
\mathrm{DM}_{1000}}{M \over M_\odot}} {3000\,\mathrm{km/s} \over V}\ 
\mathrm{y},
\end{equation}
where $V$ is the expansion velocity and $\mathrm{DM}_{1000} \equiv
\mathrm{DM_{local}}/(1000\,\text{pc\,cm}^{-3})$.  At $R = 0.1\,$ pc only $\sim
10^{-4} n_{ISM} M_\odot$ of interstellar material will have been swept up,
for an interstellar density of $n_{ISM}$ atoms/cm$^3$, so $V$ is nearly the
initial explosion velocity.  If $V$ is within the range 3000--30000 km/s of
SN ejecta then the age of the cloud $t \le T \lesssim 30$ y.  If FRB are
found within such clouds, then if repetitive bursts are observed their
dispersion measures will decrease monotonically and smoothly according to
(\ref{SNRDM}) with $R = Vt$.  The absence of known SN within the last
$\sim 30$ y at the high Galactic latitude positions of most FRB sets lower
bounds on their distances.

The hypothesis that most of the dispersion is produced in an expanding cloud
around the FRB also predicts the distribution of dispersion measures,
subject to the unknown event rate (for example, bursts may be more frequent
in younger objects, or may not occur until after a latency period from their
birth).  If we assume an age-independent event rate (at least over the
period over which the SNR provides much of the observed dispersion measure),
then from (\ref{SNRDM}) and (\ref{SNRlife}) we find
\begin{equation}
\label{dNdDM}
{dN \over d\mathrm{DM}} \propto \mathrm{DM}^{-3/2}.
\end{equation}

\begin{figure}[h!]
\centering
\includegraphics[width=4in]{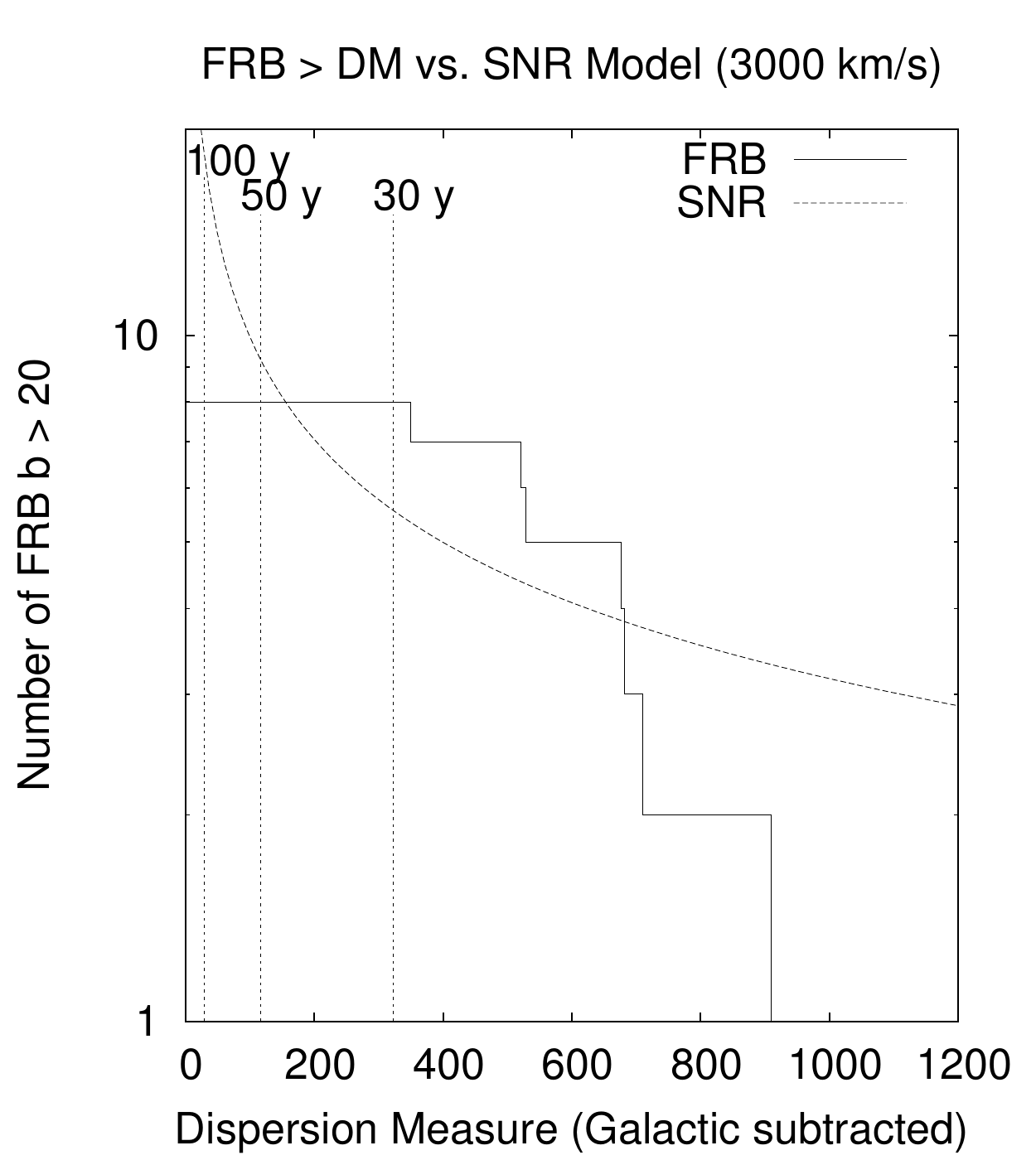}
\caption{\label{frbSNR}Cumulative distribution of dispersion measures of
high latitude ($\vert b \vert > 20^\circ$) FRB compared to a fitted
distribution of dispersion measures of expanding SNR shells.  The predicted
divergence at small dispersion measures is absent, requiring an early cutoff
to FRB activity.  The dotted lines indicate the cutoff ages required to
avoid a (not observed) divergence of the population at low DM; these ages
are given for a 1 $M_\odot$ shell with expansion speed $V = 3000$ km/s and
scale $\propto M^{1/2}/V$.  The observed FRB distribution also disagrees
with the prediction of the SNR hypothesis for $\mathrm{DM} \gtrsim 700$
pc-cm$^{-3}$.  In this model there should be no correlation between DM and
brightness because DM depends only on the SNR parameters, not the distance 
to the FRB, so a deficiency of high DM bursts cannot be explained as a
consequence of greater distance, as it might be if dispersion is
intergalactic.}
\end{figure}

In Figure \ref{frbSNR} we show the cumulative distribution of dispersion
measures of the high latitude FRB shown in the Table, and compare to the
relation $N \propto \mathrm{DM}^{-1/2}$ predicted by Eq.~\ref{dNdDM}.  Even
though only eight such FRB are known, their cumulative distribution is 
robust.  The fit is not good.

The observed distribution of DM does not display the long ``tail'' of high
DM bursts predicted by cosmologically local models in which much of the
dispersion is provided by an expanding gas cloud such as (but not
necessarily) a SNR.  Nor does it show the predicted divergence at low DM.
Even if the absence of this low-DM divergence were explained by short active
lifetimes of the FRB sources, as required by Eq.~\ref{spindown} and
consistent with Eq.~\ref{SNRlife}, the deficiency of high DM FRB argues
against the hypothesis that most of the dispersion is provided by expanding
SNR \citep{Ku14,CSP15}.  It therefore argues in favor of the cosmological
distances inferred by attributing the dispersion measures to the
intergalactic medium.  Then a deficiency of high DM FRB may be attributed to
the lesser fluence (greater luminosity distance) of more distant FRB.

This does not require rejection of the hypothesis that FRB are produced by
high field fast rotation powered pulsars.  Attributing most of the
dispersion to the intergalactic medium and accepting cosmological distances
would resolve the difficulty (Eq.~\ref{SNRlife}) that if a SNR is to provide
the dispersion its rapid expansion (especially if $V > 3000$ km/s, as found
for most SN) would require FRB activity to occur very soon after the
pulsar's birth.  However, the constraint $t < t_{spindown}$ would remain.
\subsection{Quasi-Static Clouds}
\label{Jeans}
If the dispersion occurs in a stable {\it quasi-static\/} plasma cloud, then
the condition that FRB activity occur within a time given by
Eq.~\ref{SNRlife} would no longer apply.  Such clouds need not be associated
with the compact source of the FRB, but might be, for example, dense 
formerly molecular clouds ionized by a starburst or the ionized interstellar
medium of an irregular galaxy with rapid star formation.  Specific
examples, found in regions that may be plausible sites of FRB, include dense
ionized star-forming structures \citep{Ku14} and circum-galactic nuclear gas
\citep{PC15}.

Eq.~\ref{SNRlife} is then replaced by the Jeans condition that the cloud be
stable against gravitational collapse.  This would constrain its parameters:
\begin{equation}
\sqrt{GM \over R} \lessapprox c_s = \sqrt{5 k_B T (1 + \mu_e) \over 3 m_p},
\end{equation}
where $c_s$ is the sound speed and $\mu_e \approx 0.85$ is the number of 
electrons per baryon.  Substituting $M \approx R^3 m_p n_e/\mu$ and
$\mathrm{DM} \approx n_e R$ (attributing the dispersion to the source's
plasma cloud, not the the intervening line of sight), we find
\begin{equation}
\label{RmaxJ}
R \lessapprox {5 (1 + \mu_e) \mu_e k_B T \over 3 G \mathrm{DM} m_p^2} \approx
5 \times 10^{21} {T_{8000} \over \mathrm{DM}_{1000}}\ \mathrm{cm}
\end{equation}
and
\begin{equation}
\label{neminJ}
n_e \sim {\mathrm{DM} \over R} \gtrapprox 0.6\,\mathrm{DM}_{1000}^2
T_{8000}^{-1}\ \mathrm{cm}^{-3},
\end{equation}
where we normalize the temperature $T_{8000} \equiv T/8000^{\,\circ}$K
(following \cite{Ku14}) and the dispersion measure $\mathrm{DM}_{1000}
\equiv \mathrm{DM}/$1000 pc\,cm$^{-3}$, and assume complete ionization and
cosmic abundances.  The corresponding mass 
\begin{equation}
\label{MmaxJ}
M \lessapprox {25 k_B T (1 + \mu_e)^2 \mu_e \over 9 G^2 m_p^3 \mathrm{DM}}
\approx 8 \times 10^7 {T_{8000}^2 \over \mathrm{DM}_{1000}}\ M_\odot.
\end{equation}
The hydrodynamic time
\begin{equation}
T_J \sim {R \over c_s} \lessapprox \sqrt{5 k_B T (1+\mu_e) \over 3 m_p}
{\mu_e \over G m_p \mathrm{DM}} \approx 10^8 {T_{8000}^{1/2} \over
\mathrm{DM}_{1000}}\ \mathrm{y}
\end{equation}
has no explicit dependence on the unknown parameters $n_e$, $R$ and $M$.
$T_J$ is long enough to avoid the statistical problems (Section \ref{SNR})
posed by attributing the dispersion measures to young SNR, whose
youth implies that only a very few are active with the observed dispersion
measures at any time.  The dispersive cloud could be more compact and dense
than the bounds (\ref{RmaxJ}) and (\ref{neminJ}), perhaps by a large factor.

These bounds are consistent with dense quasi-static compact clouds in the
FRB neighborhood \citep{PC15} while avoiding the rapid expansion and short
lifetime implied by attributing dispersion to young SNR.  Much smaller $R$
and $M$ and larger $n_e$ than the bounds are possible.  The bounds also
admit a protogalaxy or starburst ionized by an initial generation of hot
luminous stars, providing the observed dispersion measures.  Such sites may
be plausible locales for FRB, but give no clues to the origin of the FRB
themselves beyond indicating a relation with massive stars and high rates of
star formation and death.
\section{How Many FRB Sources?}
\label{Nsources}
Models of FRB may be divided into two general classes, those in which they
are the product of catastrophic events ({\it e.g.\/}, SN, neutron star
mergers, neutron star accretion by black holes, GRB) that destroy their
participants and cannot repeat, and non-catastrophic events ({\it e.g.\/},
giant pulsar pulses, SGR outbursts) that can repeat.  In the former case the
number of sources will equal the number of observed FRB, but in the latter
case repetitions may be observed.  The confirmed observation of a single
repetition would establish their origin in non-catastrophic events.  Here we
consider the constraints that can be placed on $N_{sources}$
non-catastrophic sources if, as at present, no repetitions have been
observed.

There are two constraints on the number of presently active detectable FRB
sources $N_{sources} \equiv BT$, where $B$ is their birth rate within the
volume from which FRB may be detected and $T$ is their active lifetime
(during which they have the defining properties of FRB, including outbursts
and dispersion measures).  If the bursts occur stochastically, without any
latency period following a burst, then the absence of coincidences among
$N_{FRB}$ observed FRB implies
\begin{equation}
\label{norep}
N_{sources} \gtrsim N_{FRB}^2 = 100.
\end{equation}
As more FRB are discovered, either coincidences (repetitions) will be
observed, implying
\begin{equation}
N_{sources} \sim {N_{FRB}^2 \over N_{coincidences}},
\end{equation}
or the lower bound of Eq.~\ref{norep} will increase.

The absence of repetitions of any individual FRB implies
\begin{equation}
\label{nsources}
N_{sources} \gtrsim \Omega_{FRB} \tau_{min} \sim 10^3,
\end{equation}
where $\Omega_{FRB}$ is the all-sky FRB rate and $\tau_{min}$ is the
empirical lower bound on the repetition time of an individual source.
\cite{KP15} and \cite{R15} estimate $\Omega_{FRB} \sim 0.03$/s with an
uncertainty as large as a factor of three and $\tau_{min} \sim 10\,$h.

If the bursts are stochastic then $\tau_{min} \gtrsim \tau_{tot}$, the {\it
total\/} time beams pointed in the {\it known\/} directions to FRB, summed
over all FRB, without observing a repetition.\footnote{It is not necessary
that a beam be pointed to a single FRB for this time because, if they all
have the same properties, staring in all directions in which an FRB has been
observed is equivalent.  It is also assumed that localization is good enough
that the chance of misidentifying a new source as a repetition of a
previously observed source is negligible; for $15^\prime$ localization and
$N_{FRB} \sim 10$ this chance is $\sim 2 \times 10^{-5}$.}  \cite{L14} found
no recurrences in $1.1 \times 10^5$ s of observations of a single FRB,
implying a 95\% confidence bound $\tau_{min} > 2.7 \times 10^4$ s, giving
the numerical estimate in (\ref{nsources}).  On the other hand, if there is
a latency period between FRB from a single source then, depending on how
observing time was distributed, $\tau_{min}$ may be as short as
$\tau_{cont}$, the longest duration of continuous observation of an
individual FRB location without a repetition.  
The conditions (\ref{norep})--(\ref{nsources}) may be used to test models
of $N_{sources}$ against the empirical parameters $N_{FRB}$, $\tau_{min}$
and $\Omega_{FRB}$, and thereby to constrain models of the sources, of their
astronomical environments, and of their distances.
\section{Distances from Comparison to SN Rate}
\label{SNrate}
Here we consider the consequences of the plausible, but unproven, assumption
that FRB are associated with SN.  The number of SNR of ages $t < T$
(Eq.~\ref{SNRlife}) associated with our Galaxy (out to distances $\sim 1$
Mpc) with sufficient column density to provide the observed FRB dispersions 
is inferred from a Galactic SN rate of 0.03/y to be $N_{SNR\ t < T} \lesssim
{\cal O}$(1), inconsistent with the observation of 10 FRB.  Eq.~\ref{norep}
provides an even stronger argument against such close, SN-associated,
sources.  Further, the all-sky FRB rate $\Omega_{FRB} \sim 0.03$/s  would
imply a repetition time of an individual source $\tau \sim N_{sources}/
\Omega_{FRB} = N_{SNR\ t < T}/ \Omega_{FRB} \sim 30\,$s.  The hypothesis of
such rapid repetitions of FRB is excluded empirically by orders of
magnitude \citep{L14,P15a}.

If FRB are associated with SN, at a rate of order one-to-one (the FRB do not
repeat), comparison of the rates of the two classes of events shows that
their distances must be cosmological:  The SN rate is estimated \citep{S07}
to be $\Omega_{SN} \approx 0.098 \times 10^{-12} M_\odot^{-1}$\,y$^{-1}$.
Standard cosmological parameters indicate a local baryon density
$\rho_{baryon} = 1.9 \times 10^{-64} M_\odot$ cm$^{-3}$ and a SN rate
$\Omega_{SN} \rho_{baryon} \approx 1.9 \times 10^{-77}$ cm$^{-3}$\,y$^{-1}$.
Comparison to the all-sky FRB rate $\Omega_{FRB} \approx 0.03\,$/s
indicates
\begin{equation}
D \sim \left({3 \over 4 \pi}{\Omega_{FRB} \over \Omega_{SN} \rho_{baryon}}
\right)^{1/3} \sim 1\ \mathrm{Gpc}.
\end{equation}
With these assumptions FRB must originate at cosmological distances, even if
much of their dispersion measures are local to their sources.  Local matter
might be the source of dispersion if FRB are giant pulsar pulses or SGR
outbursts \citep{Ku14}, but neither of these can explain the distribution
of dispersion measures (Section~\ref{DMcosmo}).

If, on the other hand, many FRB are associated with each SN, as is plausible
if they are supergiant pulsar pulses, we can still set a lower bound on the
distance out to which FRB are observed by requiring that the number of FRB
sources be at least the number of SN in the active lifetime $T_{FRB}$ of the
product of a SN:
\begin{equation}
D \gtrsim \left({3 N_{sources} \over 4 \pi \Omega_{SN} \rho_{baryon}T_{FRB}}
\right)^{1/3} \sim 10\ \mathrm{Mpc};
\end{equation}
the numerical value assumes $T_{FRB} \sim 3000$ y, the estimated active
lifetime of a SGR.  If, instead, $T_{FRB}$ is taken to be $\sim 30$ y, the
duration over which a SNR can provide the observed dispersion measure
(\ref{SNRlife}), then $D \gtrsim 50$ Mpc.  Finally, the absence of obvious
correlation with cosmologically local structure (clusters and superclusters
of galaxies) suggests $D \gtrsim {\cal O} (100)$ Mpc.
\section{Log N---Log S}
\label{logNlogS}
The distribution of fluxes or fluences for a population of objects is
a classic tool for determining their distribution in space.  It was used
in the early days of radio astronomy to exclude steady state cosmology, and
in gamma-ray burst astronomy it indicated their origin at cosmological
distances.  Because $N(S)$ is the cumulative number of sources with flux or
fluence greater than some threshold $S$, meaningful conclusions can be
drawn from remarkably small values of $N(S)$.  For FRB $S$ must be the
fluence.  In flat static (Euclidean) geometry an inverse square law holds
and $N(S) \propto S^{-3/2}$.  Results for real cosmological models cannot be
disentangled from the evolution of the FRB event rate, which is completely
unknown, so we compare to the Euclidean result, which is approximately valid
for $z \lesssim 1$.

\begin{figure}[!h]
\centering
\includegraphics[width=4in]{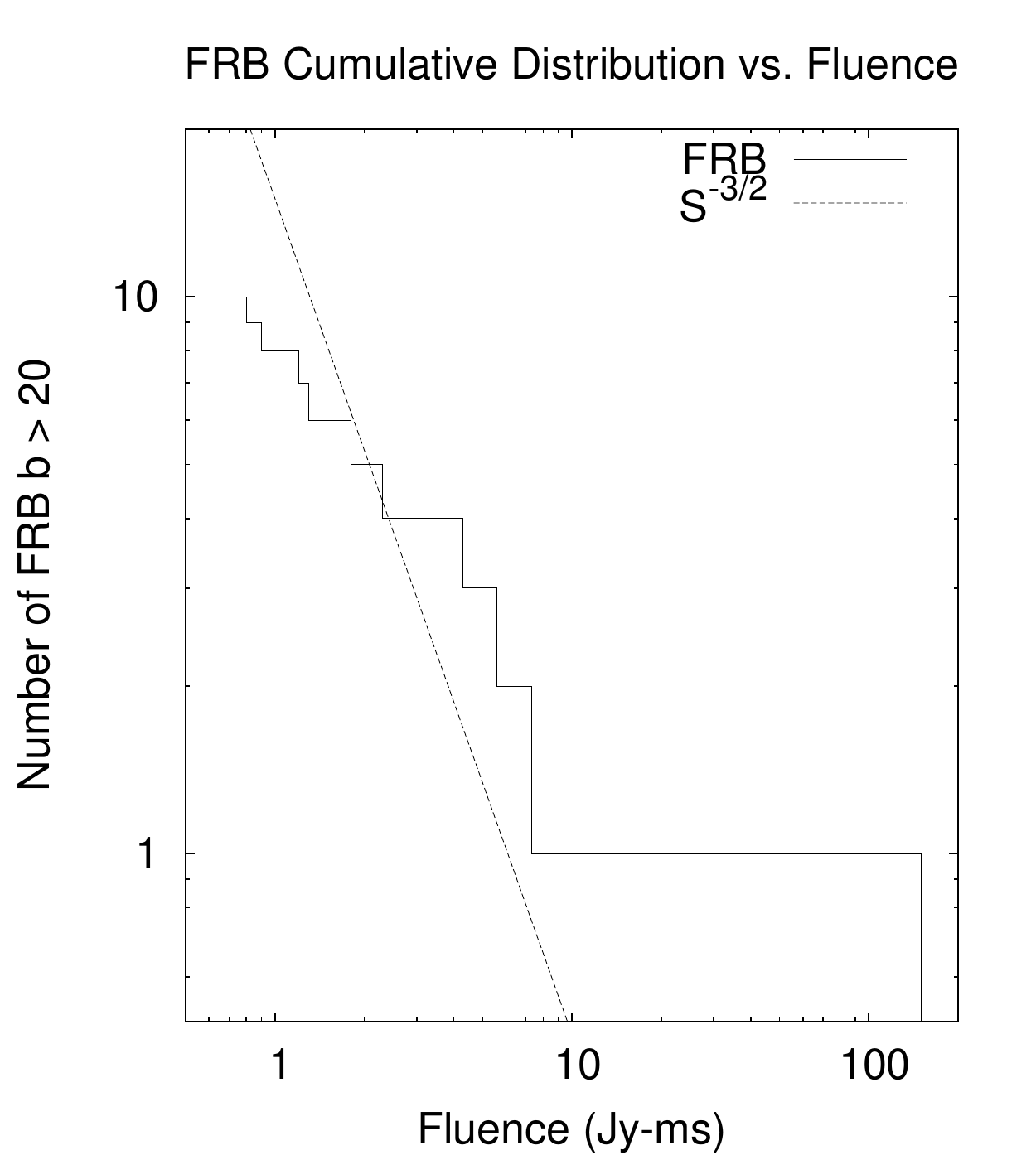}
\caption{\label{frbNvS} Cumulative $\log{N}$ {\it vs.\/}~$\log{S}$ of FRB.
The deficiency of the faintest FRB (compared to the fitted Euclidean
relation $N \propto S^{-3/2}$) may be attributable to inefficient detection
of faint FRB or to cosmological effects.  At somewhat higher fluences
agreement is good, except for the anomalous Lorimer burst estimated
\citep{L07} to have $S = 150 \,$Jy\,ms.  The absence of FRB with
$10\,\text{Jy\,ms} < S < 150\, \text{Jy\,ms}$ casts doubt on the membership of
the Lorimer burst in the FRB population.}
\end{figure}

The data are shown in Fig.~\ref{frbNvS}.  If allowance is made for the
likely incompleteness of the sample for $S < 1\,$Jy\,ms, the distribution is
consistent with homogeneous Euclidean space, as also suggested by its
isotropy, except for the anomalous Lorimer burst.  \cite{KP15} give only
a lower bound on its fluence of 31.5 Jy\,ms, rather than the specific (but
uncertain) value of 150 Jy\,ms cited by \cite{L07} and used in the Figure,
but even if this minimum is its actual value its deviation from the power
law would still be problematic.

These results do not determine a distance scale, but do indicate that FRB
are not limited to a bounded region.  Provided the deviation from the power
law at low $S$ is attributed to artefacts (for example, different research
groups or instruments having differing calibrations or thresholds for
detection or acceptance of transients as FRB), they argue against models in
which FRB are associated with bounded structures (the Solar System, Galaxy,
Local Group, Local Supercluster, local universe in which redshift and cosmic
evolution are negligible, {\it etc.\/}) outside of which the FRB density
drops.  If, on the other hand, the deviation at low $S$ is real then some
such structure is indicated.\footnote{The low fluence deviation cannot be
explained by the variation of sensitivity across a beam.  The reported
fluences are nominal values that apply if the FRB is centered in the beam,
and are not corrected for the beam pattern (a correction that cannot be made
because the location of the FRB within the beam is not known, except
possibly for the Lorimer burst that was detected in the sidelobes of two
beams in addition to its principal detection).  Each element of beam solid
angle contributes a distribution $\delta N(S) \propto S^{-3/2}$, and their
sum $N(S) \propto S^{-3/2}$.}
\section{Distribution of Dispersion Measures---Cosmological}
\label{DMcosmo}
It is possible to use the distribution of FRB dispersion measures to
constrain their spatial distribution if the dispersion is attributed to a
uniform intergalactic medium.  This is complementary to the use of the
$\log{N}$--$\log{S}$ distribution.  Unlike $\log{N}$--$\log{S}$, this 
method is specific to FRB.

The number of FRB with origin between distances of $r$ and $r + dr$
\begin{equation}
dN = K(r) r^2 dr,
\end{equation}
where $K(r)$ should include the effects of the geometry of the universe, the
redshift of the event rate, the evolution and varying density of the source
population and the K-correction (redshift of the spectrum).  The source
evolution and K-correction are unknown.  For low redshift $z \ll 1$ $K(r)
\to K$, a constant.  More generally, $K(r)$ contains three powers of $1+z$
from the variation of the source density, minus one power from the redshift
of the event rate and minus one power from the redshift of the photon energy
(because fluence rather than flux is measured the rate of photon arrival
does not contribute another factor).  If the source spectrum is a power law
$S \propto \nu^{-\gamma}$ then $K(r) \propto (1+z)^{1-\gamma}$.  With
$\gamma$ and the source evolution unknown, treating $K(r)$ as a constant may
not be far wrong for $z \lesssim 1$.

In the same spirit of approximation
\begin{align}
dr &= {c \over H_0}{dz \over 1 + z}\\
r &= {c \over H_0} \ln{(1 + z)}\\
n_e &= n_0(1 + z)^3 = 1.6 \times 10^{-7}(1 + z)^3\ \mathrm{cm}^{-3}\\
d\mathrm{DM}_{eff} &= n_e dr = {n_0 c \over H_0} (1 + z) d\ln{(1 + z)}\\
\mathrm{DM}_{eff} &= {n_0 c \over H_0} [(1 + z) - 1] \label{DM}
\end{align}
where the baryons are assumed homogeneously distributed and ionized with the
present-day density $n_0 = 1.6 \times 10^{-7}\,$cm$^{-3}$ and
$\mathrm{DM}_{eff}$ is the measured dispersion measure allowing for the
higher frequency of the observed radiation in the distant universe.  Then
\begin{equation}
{dN \over d\mathrm{DM}} = {K c^2 \over n_0 H_0^2}{\left[\ln{(1+z)}\right]^2
\over 1+z}.
\end{equation}
Using 
Eq.~\ref{DM},
\begin{equation}
{dN \over d\mathrm{DM}} = {K c^2 \over n_0 H_0^2}{\left[\ln{\left(
1 + {H_0 \mathrm{DM} \over n_0 c}\right)}\right]^2 \over 1 + {H_0
\mathrm{DM} \over n_0 c}}.
\end{equation}
This is readily integrated
\begin{equation}
\label{NvsDM}
N = {K c^3 \over 3 H_0^3}\left[\ln{\left(1 + {H_0 \mathrm{DM} \over
n_0 c}\right)}\right]^3 \propto \left[\ln{\left(1 + {\mathrm{DM} \over
682\,\text{pc\,cm}^{-3}}\right)}\right]^3.
\end{equation}

\begin{figure}[!h]
\centering
\includegraphics[width=4in]{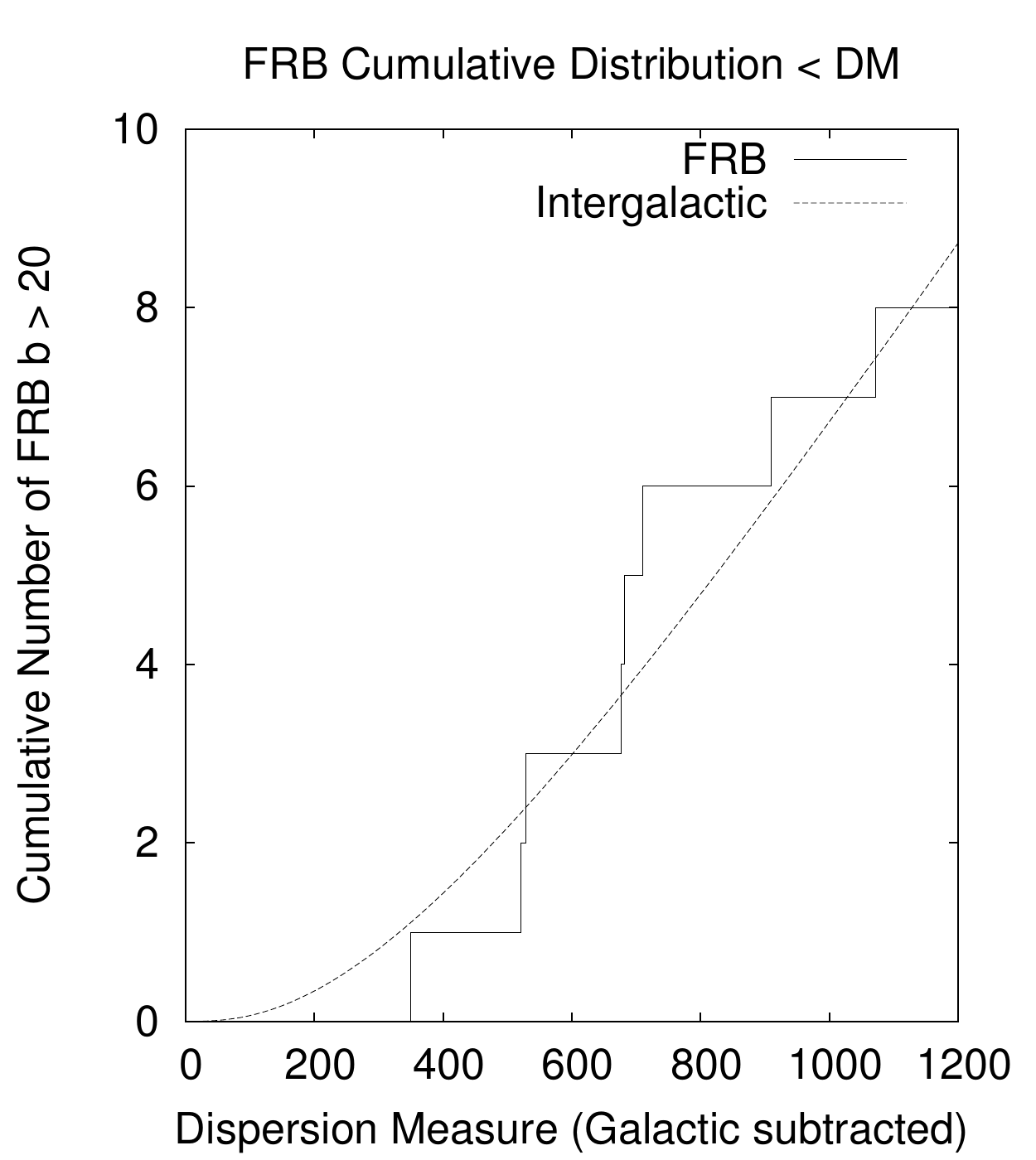}
\caption{\label{frbcosmo}The cumulative distribution of FRB (the eight at
high Galactic latitudes for which the Galactic contributions to their
dispersion measures are small and can be reliably subtracted) {\it vs.\/}
the extraGalactic part of their dispersion measures.  The ``Intergalactic'' 
curve (Eq.~\ref{NvsDM}) describes a simple model of dispersion attributable
to the intergalactic medium.  Agreement is good.}
\end{figure}

Figure~\ref{frbcosmo} shows the cumulative distribution of $N$ as a function
of DM and a curve of the form of Eq.~\ref{NvsDM} with one fitted parameter
($K$).  The fit is good.  The most striking feature is the absence, in both
the data and the theoretical curve, of FRB with low dispersion measures.
This is explained by the small volume $\propto z^3$ of space with low $z$;
Euclidean geometry and the absence of cosmic evolution are good
approximations at small redshifts.  This supports the interpretation of the
dispersion measures of FRB as resulting from propagation through
intergalactic space, independent of any specific model of their origin.
In contrast, models \citep{Ku14,PC15} that attribute FRB dispersion measures
to quasi-static clouds (Section~\ref{Jeans}) in their vicinities make no
specific predictions of the distribution of DM, but do not naturally explain
the data shown in Fig.~\ref{frbcosmo}.

This result is robust against changes in the assumptions.  For example,
ignoring the $(1+z)^{-2}$ factor in $d\mathrm{DM}_{eff}$ that results from
the from the $z$ dependence of the frequency of radiation along its path
leads to a fit similar to that of Figure~\ref{frbcosmo} (but with a
different $K$).  This robustness is not surprising, because the model is
dominated by the $\propto z^3$ dependence of the volume and $N$ in the local
(Euclidean) limit.  

\section{Discussion}
\label{discuss}
Understanding of the high brightness emission of FRB may be as elusive as
understanding of the emission of radio pulsars has been, nearly 50 years
after their discovery.  Radio pulsars were very quickly identified on the
basis of their periods and slowing with rotating neutron stars.  No such
clues exist for FRB.

Even with only ten published FRB, and without a specific model of their
sources, statistical arguments can constrain their origin, both physical and
spatial.  Their qualitative resemblance to radio pulsars, and in particular
to the rare giant pulses of a few pulsars, suggests that they may be
analogous events, writ large.  Elementary energetic considerations show that
this may be possible, but if so their sources must be very young neutron
stars.

This leads to the hypothesis that, surrounded by young, dense, compact SNR,
much of the dispersion measures of FRB may be attributed to those SNR
\citep{Ku14,CSP15}.  If so, the distances to FRB may be much closer than
those indicated by attributing their dispersion measures to the
intergalactic medium, relaxing the constraints on their energetics implied
by the assumption of cosmological distances.  This hypothesis predicts a
specific form for the distribution of FRB dispersion measures.
Unfortunately, comparison to the data (Fig.~\ref{frbSNR}) shows a poor fit
at both low and high dispersion measures.  The former might be explained by
termination of FRB activity as the neutron star spins down (or ages in some
other manner), in analogy to the pulsar ``death line'', but the absence of
FRB with very high dispersion measures is an argument against this origin of
their dispersion.

The classic $\log{N}$--$\log{S}$ test, where $S$ is the FRB fluence rather
than the (unmeasured) flux, may also be applied (Fig.~\ref{frbNvS}).  A fit
to the relation $N \propto S^{-3/2}$ applicable to Euclidean geometry
(cosmologically local sources) is adequate at intermediate values of $S$.  A
deficiency of low fluence FRB may be explained by sample incompleteness.  If
so, the fit at higher fluences indicates a source population with $z
\lesssim 1$, but excludes a spatially bounded distribution such as the
Galactic halo.

Similar conclusions may be drawn from the distribution of FRB dispersion
measures less than a cutoff (Fig.~\ref{frbcosmo}).  A simple model that
includes both cosmologically local and distant sources (but that ignores the
unknown evolution of the source population, and hence must be considered
only qualitative for $z \gtrsim 1$) provides a good fit.  In contrast
to the similar inferences from the $\log{N}$--$\log{S}$ data, this conclusion
does not depend on additional explanations of apparent deviations from the
model.  A robust conclusion, independent of any caveats applicable for $z
\gtrsim 1$, is that the source population is homogeneously distributed in
space at smaller $z$.  This is an essentially Euclidean conclusion, but the
satisfactory fit for $z \sim 1$ suggests that the distances are those
inferred from the assumption that the dispersion measures are intergalactic.
This is supported by the failure of the alternative expanding cloud (SNR)
models of the dispersing matter (Fig.~\ref{frbSNR}).  Thee fact that this
cosmological model of the distribution of dispersion measures is consistent
with the data implies that nearly all the dispersion is intergalactic, in
contrast with models \citep{Ku14,Ku15} in which a large portion of the
dispersion measure is attributed to gas local to the FRB.

The extraordinarily intense Lorimer burst is not consistent with $N \propto
S^{-3/2}$.  The absence of weaker events was noted by \cite{L07}, and
indicates that it comes from a different population.  With a fluence
$\sim 100$ times the detection threshold, a homogeneous distribution in
Euclidean space (its comparatively small dispersion measure indicates it
must be cosmologically local) would suggest that $\sim 1000$ weaker events
should have been detected, in contrast to the 10 actually detected, an
anomaly that is significant at the $\approx 99$\% level (the detection
of 10 weaker events implies a 1\% probability of detecting one as strong as
the Lorimer burst).  If this was neither a statistical fluke (because of the
need to reject anthropogenic transients, early searches may have had
effective detection thresholds higher than their nominal thresholds) nor
anthropogenic interference, it implies the Lorimer burst came from a
distinct space-limited population whose local density far exceeds its mean
cosmological density.  Speculative possibilities include classes of
intrinsically less luminous bursts in the Galactic halo or the Local
Supercluster.
\appendix
\section{Space Radar as a Source of Chirped Pulses?}
\label{chirp}
Since the discovery of FRB there has been concern that they might be
anthropogenic, rather than astronomical, phenomena.  This concern was
exacerbated by the discovery by \cite{BS11} of manifestly anthropogenic
perytons, whose sources were recently identified by \cite{P15b} as
microwave ovens.  Could FRB also be anthropogenic interference, perhaps
chirped radar pulses?  A ground-based radar might enter of sidelobes of a
radio telescope, but would appear in all 13 Parkes beams, unlike the FRB.
Hence we consider a radar on a satellite that might pass through a beam.

Radar systems may use chirped emission, compressed upon reception into
narrow pulses, in order to obtain accurate range measurements without
requiring excessive peak transmitted powers.  The observation of FRB in a
single beam at Parkes, in contrast to perytons \citep{BS11}, indicates a
distance $\gtrsim 20$ km, outside the first Fresnel zone, consistent with
a radar satellite.  There is no obvious reason for a radar to have a chirp
$\omega \propto t^{-1/2}$ as observed, nor is there obvious reason not.
However, the observed dispersed pulse durations of several tenths of a
second would imply, for monostatic radar, target distances of at least half
that light travel time to avoid interference of the transmission with the
received scattered radiation.  Such distances would be $\sim 10^{10}\,$cm,
far beyond the range to plausible targets, and the return would be
undetectably weak.

In contrast, bistatic radar can use arbitrarily long pulses.  The pulse
repetition intervals would have to have been longer than the lengths of time
the radars were anywhere in the 13 beams of the Parkes Multibeam Pulsar
Survey (about 0.3 s for a radar near the zenith in low Earth orbit), yet the
pulse durations must have been shorter than the time required to cross a
single beam.  This explanation would also require at least as many radar
satellites, each with a different chirp rate, as FRB, because each FRB had a
different dispersion measure, or satellites whose chirp rates were variable
in some non-obvious manner.  This combination of requirements makes the
hypothesis of interference by an orbital chirped source implausible.

I thank J. Goodman and T. Piran for useful discussions, and an anonymous
referee for pointing out the invalidity of an essential assumption of an
earlier version of this paper.
\appendix

\end{document}